\documentclass{elsart5p}

\usepackage{natbib}

\usepackage{graphics}
 \usepackage{graphicx}
\usepackage{epsfig}

\usepackage{amssymb}

\begin{document}

\begin{frontmatter}



\title{INTEGRAL high energy sky: The keV to MeV cosmic sources}

%

\author[rome]{Pietro Ubertini,}
\author[rome]{Alessandra De Rosa,}
\author[rome]{Angela Bazzano,}
\author[bologna]{Loredana Bassani,}
 \author[bologna]{Vito Sguera}
 \author[]{on behalf of the INTEGRAL survey team}
\address[rome]{INAF-IASF/Roma Via del Fosso del Cavaliere, 100, 00133 Roma, Italy}
\address[bologna]{INAF-IASF/Bologna, Via Gobetti 101, Bologna, Italy} 

\begin{abstract}
After almost 5  years of operation, ESA's International Gamma-Ray Astrophysics Laboratory (INTEGRAL) Space Observatory has unveiled a new soft Gamma ray sky and produced a remarkable harvest of results, ranging from identification of new high energy sources, to the discovery  of dozens of variable sources to the mapping of the Aluminum emission from the Galaxy Plane to the presence of electrons and positrons generating the annihilation line in the Galaxy central radian. INTEGRAL is continuing the deep observations of the Galactic Plane and of the whole sky in  the soft Gamma ray range. The new IBIS gamma ray catalogue contains more than 420 sources
detected above 20 keV.
We present a view of the INTEGRAL high energy sky with particular regard to sources emitting at high energy, including Active Galactic Nuclei (AGN), HESS/MAGIC counterparts and new view of the cosmic gamma ray diffuse background.

\end{abstract}


\end{frontmatter}

\section{Introduction}
\label{intro}

Following almost 5 years of successful operations, INTEGRAL has significantly changed our vision
of the Universe through its observations of the gamma-ray sky. The telescopes aboard the satellite
have revealed hundreds of sources of different types and new classes of objects. INTEGRAL is providing surveys of the hard X-ray and soft gamma-ray sky, with a census of the source populations and first-ever all sky maps in this so far unexplored energy range. We present here the new vision of the high energy sky as painted by INTEGRAL observatory.
The main characteristics of INTEGRAL are described in Sect. \ref{integral}, while in Sect. \ref{catalogue} we report on the results of the 3rd survey. In Sects. \ref{SGXB}, \ref{agn}, \ref{hess}  we described the new results derived by INTEGRAL observations about Supergiant High Mass X-ray Binaries (SGXBs), Active Galactic Nuclei (AGN) and Very High Energy (VHE) TeV sources.

\begin{figure*}[t!]
\vspace{-1cm}
\includegraphics*[height=0.75\textheight]{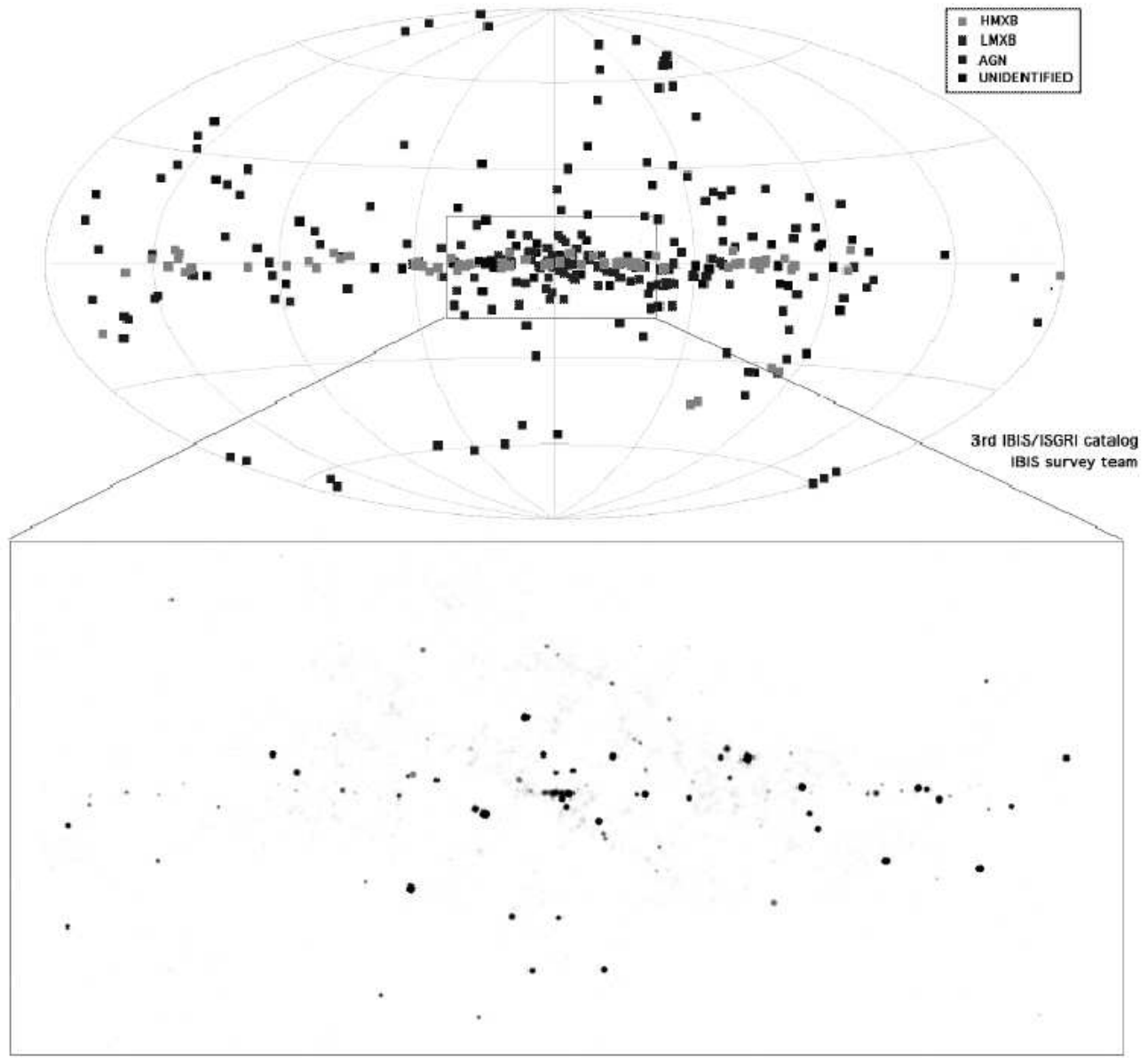}
\caption{The upper image shows the distribution on the sky of four of the main soft gamma-ray source populations observed in the third INTEGRAL/IBIS survey catalogue. This newly-released catalog contains 421 sources. Of the known systems, the low-mass X-ray binaries (LMXB) are old systems mainly populating the galactic bulge, the high-mass X-ray binaries (HMXB) are younger systems seen along the galactic plane, and the active galactic nuclei (AGN) are extragalactic sources seen over the whole sky. Around one in four of the sources seen by INTEGRAL are unidentified, and their distribution is also shown. 
The lower image shows a false colour image of the central region of our galaxy. This is a composite image based on all-sky IBIS/ISGRI maps in three energy windows between 17 and 100 keV and represents the true 'X-ray colours' of the sources. Grey sources are dominated by emission below 30 keV, while black sources have harder spectra, emitting strongly above 40 keV (Credit: IBIS Survey Team)
}
\label{fig:survey}
\end{figure*}

\section{The INTEGRAL Observatory}
\label{integral} 
The ESA INTEGRAL (International Gamma-Ray Astrophysics Laboratory) observatory was
selected in June 1993 as the next medium-size scientific mission within ESA Horizon 2000 programme.
INTEGRAL \citep{winkler03} is dedicated to the fine spectroscopy (2.5 keV FWHM @ 1 MeV) and fine imaging (angular resolution: 12 arcmin FWHM) of celestial gamma-ray sources in the energy range 15 keV to 10 MeV. While the imaging is carried out by the imager IBIS \cite{ube03}, the fine spectroscopy if performed by the spectrometer SPI \cite{vedrenne03} and  coaxial
monitoring capability in  the X-ray (3-35 keV) and optical (V-band, 550 nm) energy ranges in provided by the JEM X and OMC instruments \cite{lund03,mass03}.   
SPI, IBIS and Jem-X, the spectrometer, imager and X-ray monitor are based on the use of coded aperture mask technique thath is a key feature to prove images at energy above tens of  KeV, where photons focussing become impossible using standard grazing technique. 

Moreover, coded mask feature the best background subtraction capability because
of the possibility to observe at the same time the Source and the Off-Source sky. This is achieved at the same time for all the sources present in the telescope field of view.
In fact, for any particular source direction, the detector pixels are split into two sets: those capable of viewing the source and those for which the flux is blocked by opaque tungsten mask elements.  
This very well established technique is discussed in detail by \cite{skiner03} and is extremely effective in controlling the systematic error in all the telescope observation,   working remarkably well for weak extragalactic field as well for crowded galactic regions, such as our Galaxy Center.
The mission was conceived since the beginning as an observatory led
by ESA with contributions from Russia (PROTON launcher)
and NASA (Deep Space Network ground station).

INTEGRAL was launched from Baikonur on Oct.17th, 2002 and  inserted into an highly eccentric orbit  (characterized by 9000 km perigee and 154000 km apogee).  The high perigee in order to provide long uninterrupted observations with nearly constant background and away from the electron and proton radiation belts. Scientific observations can then be carried out while the satellite being above a nominal altitude of 60000 km, while entering the radiation belts, and above 40000 km, while leaving them.

This strategic choice ensure about  90\% of the time is used for scientific observations with a data rate of realtime, 108 kbps science telemetry received from the ESA station of
Redu. 
The data are received by the INTEGRAL Mission Operation Centre (MOC) in Darmstadt (Germany) and relayed to the Science Data Center
(ISDC)  \cite{courv03} which provide the final consolidated data products to the observes and later archived for public use. The proprietary data become public one year after distribution to single observation PIs.

\section{The sources catalogue} 
\label{catalogue} 

The 3rd INTEGRAL sources catalogue \cite{bird07}, compiled using all the IBIS data available up the and of May 2006 (40 Ms exposure time), encompasses 421 sources, detected above 4.5 $\sigma$ in the energy range 18-100 keV; in Figure \ref{fig:survey} we show the  distribution on the sky of four of the main soft gamma-ray source populations.
It does not come as a surprise that the majority of the IBIS sources is located at low galactic latitudes, in fact INTEGRAL is frequently observing in the Galactic Plane and the sky coverage is far for uniform (see fig.1 of \cite{bird07}).  Irrespective of source location, the identification process is based on a multiwavelength approach, taking advantage of Radio, IR and X-ray archival data. Ad hoc optical and IR observing campaigns are also actively pursued \cite{masetti06}. 
Considering only the firm identifications, 171  sources (i.e. 41\%)  have been associated with galactic accreting systems, 122 (i.e. 29\%) with  extragalactic objects, 15 with different classes of celestial emitters, while 113 (i.e. 26\%) are still awaiting identification.  
The galactic identifications are divided into 21 CV systems (9 of which are new detections with emission extending up to 100keV), 65 high-mass X-ray binaries (HMXB) and 78 low-mass X-ray binaries (LMHB). Even if  INTEGRAL continues to detect LMXB, the rate of discovery is much lower than for the high mass systems. In particular, the HMXB sample encompasses also 19 new INTEGRAL sources which have been identified with Be binary systems on the basis of their spectral characteristic and/or transient behavior, pointing out the emergence of a new class of supergiant fast X-ray transient (SFXT). The efforts to identify new INTEGRAL sources (known as IGRxxxx.yyyy) resulted in 8 firm identifications and 4 probable ones and we will discuss these objects in Sect. \ref{SGXB}.

\begin{figure*}[t]
\begin{minipage}{4cm}
 \includegraphics[height=.2\textheight]{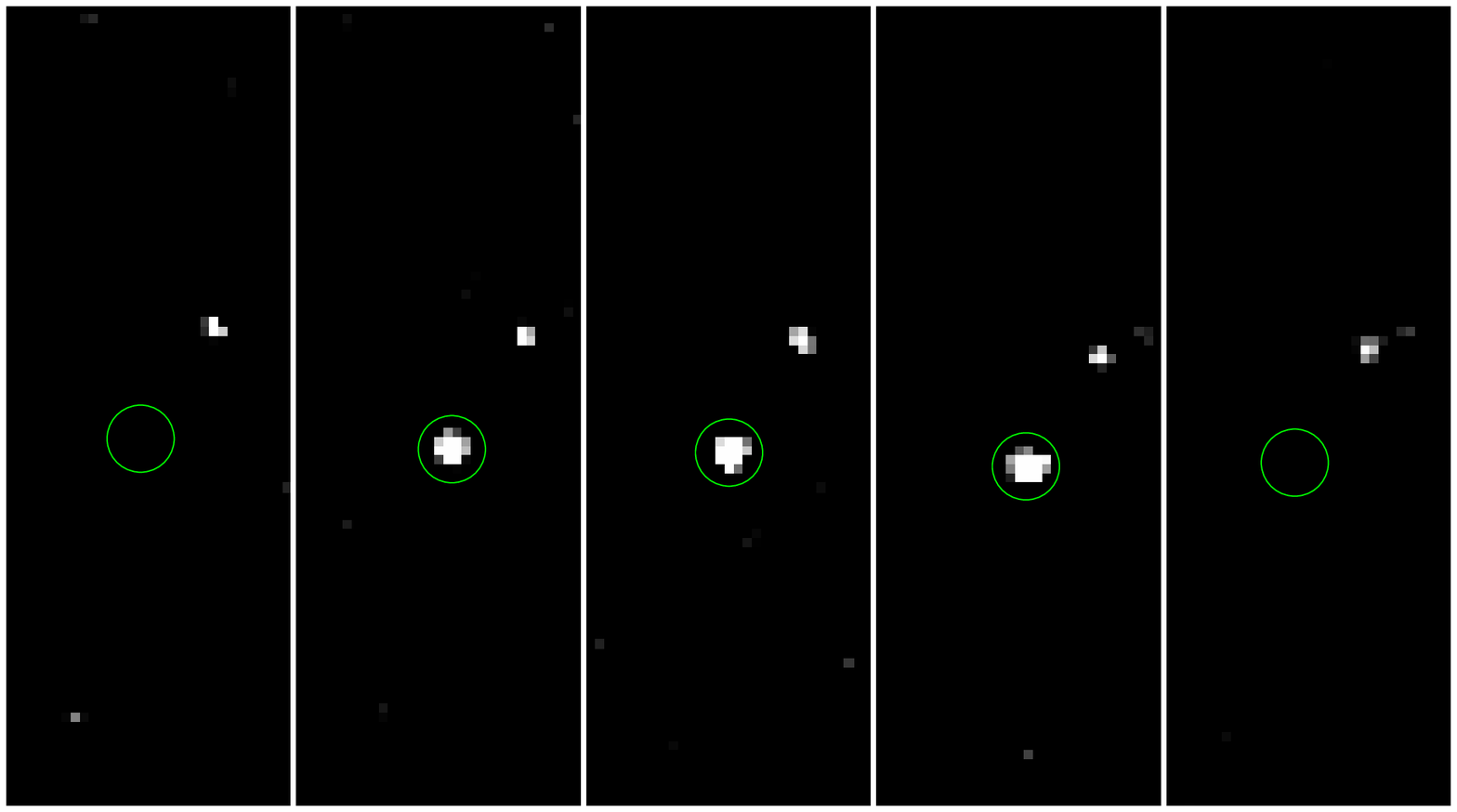}
\end{minipage}
\hspace{5cm}
\begin{minipage}{4cm}
\includegraphics[height=.4\textheight,angle=-90]{XTE1739.ps}
\end{minipage}
\caption{Left: ISGRI Science Window (ScW) image sequence (20--30~keV) of a fast X-ray outburst of
XTE~J1739$-$302 (encircled). 
The duration of each ScW is $\sim$ 2000 s.
The source was not detected in the first ScW (significance less than 2$\sigma$), then
it was detected during the next 3 ScWs with a significance, from left to right, equal to 14$\sigma$,
16$\sigma$ and 23$\sigma$, respectively.
Finally in the last ScW the source was not detected (significance less than 2$\sigma$).
A weak persistent source (1E 1740.7-2942) is also visible in the field of view. Right: ISGRI light curve (20--60 keV) of the fast X-ray outburst from XTE~J1739$-$302 shown in figure
1.}
\label{fig:sgxb}
\end{figure*}

\subsection{The sky above 100 keV} 

The high energy sky population has been obtained by INTEGRAL at energies
above 100 keV and the result discussed in  \cite{bazzano06} that have
provided the first catalogue IBIS sources characterised by very hard
spectral components. The catalogue provides evidence for  49 sources
detected with a significance above 4  in the 100-150 keV energy range of
which only 14 seen above 150 keV.
Black Holes and accreting Neutron Stars systems mainly populate the high
energy sky. The high energy emitting objects have been firmly identified
with  26 LMXBs, 6 HMXBs , 2 anomalous X-ray pulsars, while 1 object has
still a non clear counterpart and is perhaps a Black Hole candidate in a
LMXB. Among the others 1 Soft Gamma Repeater, 3 isolated pulsars and 10
AGNs have been detected.  In the higher energy range ($>$150 keV) the main
emitting sources are Black Hole candidates/microQSOs. Finally, regarding
spectral characteristics,  the softest one is ScoX-1, even if showing a
transient very hard X-Ray tail extending up to 500 keV \cite{disalvo06} while the two hardest are the 2 anomalous X-ray pulsars.

\section{Supergiant High Mass X-ray Binaries}
\label{SGXB}

Supergiant High Mass X-ray Binaries (SGXBs) are systems composed of an accreting compact object
(magnetized neutron star or black hole) and 
a massive supergiant early-type star (OB). The X-ray emission is powered by accretion of material
originating from the donor star through strong 
stellar wind or occasionally by Roche-lobe overflow. 
Up to recently SGXBs were believed to be very rare objects due to the evolutionary timescales involved;
supergiant stars have a very short lifetime. 
This idea was supported by the fact that only  a dozen of SGXBs have been discovered in almost 40 years of
X-ray astronomy
and it was largely believed that the dozen of known objects represented a substantial fraction of all SGXBs in  our Galaxy. They are 
known to be bright persistent X-ray sources rather stable in the long run with  X-ray luminosities  in the
range 10$^{36}$--10$^{38}$ erg s$^{-1}$,
depending on the accretion mode.

The INTEGRAL satellite is changing  this  classical picture on SGXBs.
Since its launch in 2002, INTEGRAL in just a few years doubled the population of SGXBs  discovering about
17 new systems. 
The majority of them (13) are persistent X-ray sources which escaped previous detection because of their
very strong absorbed nature, being the  N$_H$ 
typically greater than 10$^{23}$ cm$^{-2}$ \cite{walter06, chaty06}.
As well as the highly absorbed persistent SGXBs, INTEGRAL is also  playing a key role in unveiling another
kind of SGXBs, not strongly absorbed, which escaped detection by previous X-ray missions mainly because of
their fast X-ray transient behaviour, a characteristic never seen before. They have been labeled as Supergiant  Fast X-ray Transients SFXTs \cite{negu05,sguera05,sguera06}
because spend most of the time in quiescence  with X-ray luminosity values 
or upper limits in the range 10$^{32}$--10$^{33}$  erg s$^{-1}$ and then occasionally  undergo fast X-ray
transient 
activity lasting typically few hours, rarely few days, reaching peak-luminosities in the range 
10$^{36}$ - 10$^{37}$ erg s$^{-1}$.  Their outbursts show complex structures characterized 
by several fast flares with both rise and decay times of typically a few tens of minutes. This kind of X-ray behaviour is very surprising since SGXBs were
seen up to recently only as bright
persistent  X-ray sources. XTE J1739$-$302 is the best studied of these new systems \cite{smith06,negu06,sguera05,sguera06}. 
Figure \ref{fig:sgxb} show the ISGRI Science Window  significance map sequence in 20--30 keV energy range (left panel) and the correspondent
ISGRI 
light curve in 20--60 keV  energy range (right panel) of a typical outburst from XTE J1739$-$302 detected by INTEGRAL.

SFXTs are difficult to detect because of their very transitory nature.
The IBIS instrument  on board the INTEGRAL satellite  is 
particularly suited to the detection of new or already known SFXTs thanks to its large FOV, good
sensitivity, good angular resolution and point source location accuracy, continuos monitoring of the
galactic plane.
To date, 8 SFXTs have been reported in the literature  in just a few years, this is a huge achievement  if
we take into account that their number is almost comparable to that of classical persistent SGXBs
discovered in 40 years of X-ray astronomy.
Four out of eight firm SFXTs are new hard X-ray sources discovered by INTEGRAL while the remaining were
discoverd by previous X-ray satellites (ASCA, RXTE, BeppoSAX WFCs) however INTEGRAL detected several fast
X-ray outbursts unveiling or strongly confirming their fast X-ray transient nature. 
Moreover, there are a few more optically unidentified X-ray sources which 
display a fast X-ray transient behaviour strongly resembling that of firm SFXTs and so they are candidate
SFXT 
(\cite{negu05,sguera06}). 

The physical reasons of the fast X-ray outbursts displayed by SFXTs is still unknown. 
It has been suggested that the origin is not related to the compact object but it must be related to the
early-type
supergiant donor star. It could be that the wind accretion mass transfer mode from the supergiant star to
the compact object in SFXTs is 
different from  that in classical persistent SGXBs. The supergiant star probably ejects material in a
non-continuous way, 
its massive wind could be highly inhomogeneous, structured, characterized by a clumpy nature and the
capture of these clumps
by  a nearby compact object could then produce fast X-ray flares (\cite{negu05,intzand05}).
As for the  low  quiescent X-ray luminosities of SFXTs, this could be explained by very eccentric orbits
and long orbital periods much greater than those of classical persistent SGXBs which are typically in the
range  1-14 days. Because of this, the compact object in SFXTs spends most of the 
time far away from the supergiant donor star \cite{negu05}. This 
should implies a periodicity of the fast outbursts as they should occur always relatively close to the
periastron passage but to date no periodicity has been observed from all but one firm SFXTs. IGR
J11215$-$5952 is the only SFXTs
for which an orbital period of $\sim$ 330 days has been reported by \cite{sidoli06}.
A complete and definitive  explanation  of the physical reasons responsible for the very unusual fast
X-ray transient behaviour
of SFXTs can only be possible through a knowledge of their  orbital parameters and system geometry.

In the light of these new and exciting INTEGRAL results, the size of the population of SGXBs in our Galaxy
could have been severely underestimated. 
In particular, the class of SFXTs could be much larger than the 8 firm sources reported in the literature.
An entire population of still undetected SFXTs could be hidden in our Galaxy. Ongoing observations with
INTEGRAL may yield many newly discovered SFXTs as well as provide breakthrough information to further
insight into the system geometry, so allowing the study of the physical 
reasons behind their very unusual fast X-ray transient behaviour.

\section{The extragalactic sky seen by IBIS: from the nearest to the farthest AGN}
\label{agn}

\begin{figure}
\hspace{-0.8cm}
\begin{minipage}{4 cm}
\includegraphics[height=.4\textheight]{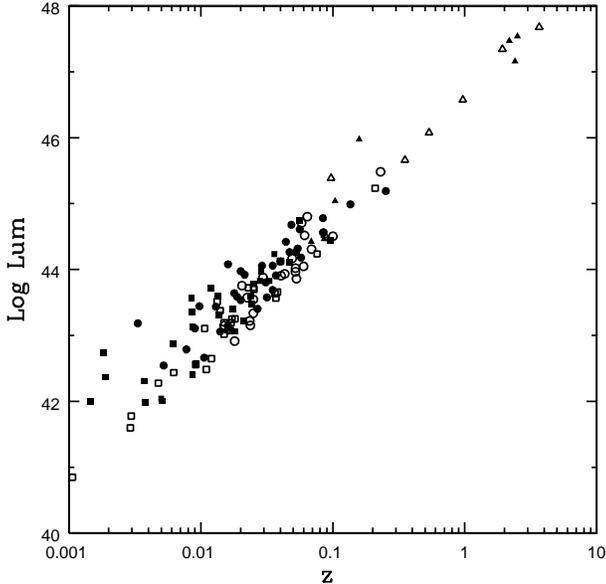}
\end{minipage}
\caption{Redshift versus 20-100 keV Luminosity for our sample of IBIS detected AGN, which have been optically classified}
\label{L_z}
\end{figure}

The third IBIS survey \cite{bird07} has provided a significant improvement in the detection of extragalactic
objects due to the larger sky coverage available. Using a number of recent identification/classification 
being provided through optical spectroscopy and catalogue searches, we now have a list of 
128 secure AGN and around 20 candidates.  
For those objects with known distance, we plot in Figure \ref{L_z} the  20-100 keV luminosity against redshift,
to show the large range in these parameters sampled by our survey. From this figure it 
is also possible to estimate  our sensitivity limit which is around 
1.5$\times$10$^{-11}$ erg cm$^{-2}$s$^{-1}$.

\begin{figure}
\hspace{-0.8cm}
\begin{minipage}{4 cm}
\includegraphics[height=.4\textheight,angle=-90]{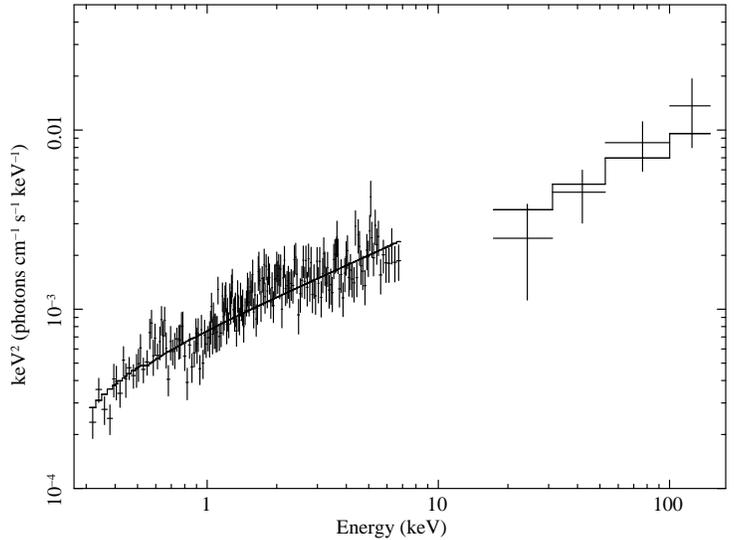}
\end{minipage}
\caption  {Broad band spectrum of 
NGC4395 fitted with a simple power law (continuous line in figure): stacked XRT data
to the left and IBIS data to the right} 
\label{4395}
\end{figure}

Within the  sample of optically classified objects, 114 objects are  Seyfert galaxies and 13 are blazars; within the Seyfert sample 58 objects are AGN  of type 1-1.5 while 57 are of type 2, i.e. a ratio 1:1,  which 
illustrates the power
of gamma-ray surveys to find narrow line AGN.  About 10$\%$  of our sample is made of radio loud AGNs.
The range of redshift probed by  our sample is 0.001-3.668 while the 20-100 keV luminosities span from
$\times$ 10$^{41}$ erg s$^{-1}$ to $\times$ 10$^{48}$ erg s$^{-1}$.\\
The closest object detected is the Seyfert 1.8 galaxy NGC4395 (z=0.001)
while the farthest one is IGR J22517+2218 (z=3.668).
NGC 4395 holds the distinction of hosting the
Seyfert 1 nucleus with the lowest known optical luminosity \cite{filippenko89}, and one of the intrinsically weakest nuclear X-ray sources
observed so far \cite{cappi06}. NGC 4395 is also
highly unusual because it harbours a small black hole mass of 10$^{4-5}$ solar masses compared to
typical values found in more typical Seyfert galaxies \cite{panessa06}
In addition, it shows some of the strongest X-ray variability known in radio-quiet AGN \cite{vaughan05} with variations of nearly an order of magnitude in  a few thousand seconds. In other respects, however, such as optical and ultraviolet emission (and absorption) line properties, it resembles its more luminous counterparts \cite{filippenko89}. This source has never been reported at
energies above 10 keV and so the INTEGRAL detection is the first one in these high energy band. NGC4395 is detected by IBIS
with a significance of $\sim$5$\sigma$; a simple power law provides a
good fit to the IBIS data in the range 17-150 keV ($\chi^2$=8.8 for 10 d.o.f.):
the  photon index $\Gamma$ is flat (1.0$\pm$0.7) and the  20-100 keV 
flux is 1.4 $\times$ 10$^{-11}$ erg cm$^{-2}$ s$^{-1}$. 
To construct the broad band spectrum of the source we have used  recent X-ray data obtained  by XRT \cite{hill04} on board Swift \cite{gehrels04}. We have analysed a set of 4 observations having  the longer exposures and then combined them together to obtain an average spectrum of the source in the 2-10 keV band which we used in conjunction  with the IBIS data.
This combined X-ray/gamma-ray spectrum is shown in Figure \ref{4395}: it is well described by a flat power law with photon index $\Gamma$=1.4$\pm$0.4 ($\chi^2$=256.2 for 203 d.o.f.); the cross calibration constant is 0.74$\pm$0.24 which indicates a good match between the two sets of data despite the strong variability of the source. The observed
the 2-10 keV flux is 5 $\times$ 10$^{-12}$ erg cm$^{-2}$ s$^{-1}$ similar to what generally observed by revious X-ray instruments \cite{moran05}.  No sign of absorption is evident in the XRT data but this maybe related to the poor statistical quality of the spectra used; if we force the spectrum to have an
absorbed component in the range 1-5 $\times$10$^{22}$ cm$^{-2}$ as previously observed, then the spectrum steepen to a $\Gamma$ value greater than 2.7. The flat spectrum observed by IBIS is clearly at odds with the typical high energy spectra of more classical Seyfert galaxies; clearly more data are needed to confirm this observational evidence and so NGC4395 is a source to keep looking at. 

\begin{figure*}[t]
\begin{minipage}{8cm}
\includegraphics[height=.3\textheight]{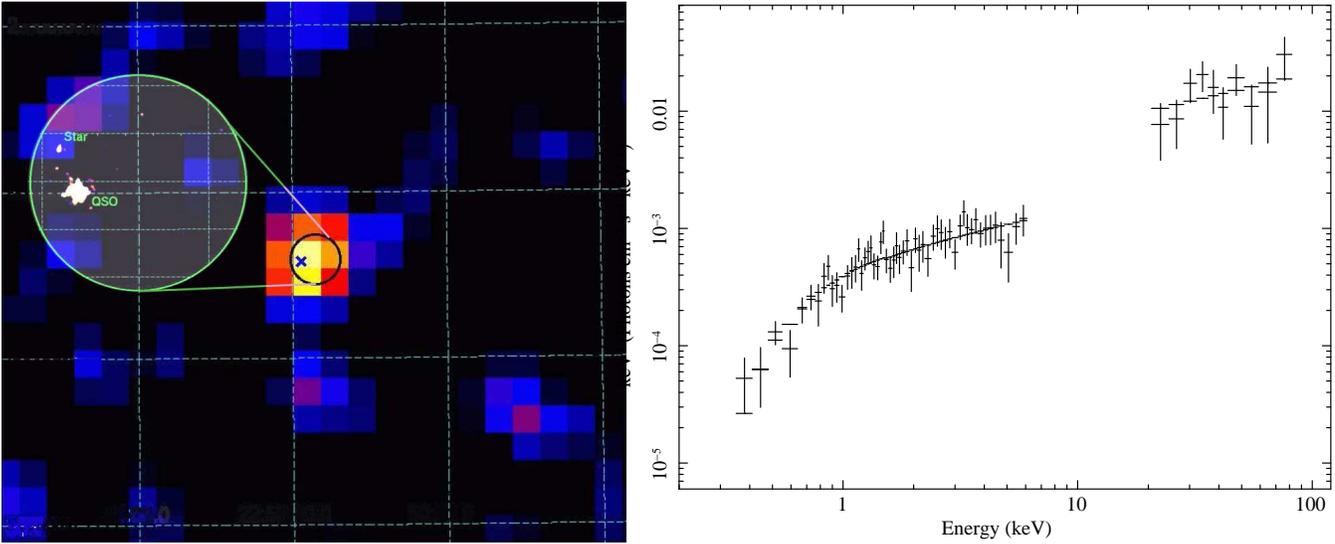} 
\end{minipage}
\begin{minipage}{8cm}
\includegraphics[height=.4\textheight,angle=-90]{hig_z_qso_broadspec.ps}
\end{minipage}
\caption {Left: IBIS/ISGRI 20-100 keV image of  IGR J22517+2218. 
The circle corresponds to the INTEGRAL/IBIS error box while the X is the optical position of MG3 J225155+2217.
The zoom to the left  is insteadthe XRT image of the IBIS error box which shows the detection of only one bright X-ray source identified with the high z QSO MG3 J225155+2217; the other weaker but still visible source is a star with coronal emission detected only below 3 keV. Right: Broad band spectrum of 
IGR J22517+2218/MG3 J225155+2217 fitted with an intrinsically absorbed power law (continuous line in figure): stacked XRT data
to the left and IBIS data to the right}
\label{qso}
\end{figure*}

On the other side of the redshift range, we have a newly identified IBIS source which has been recently associated to a very high z quasar \cite{bassani07}. IGR J22517+2218 was detected by IBIS with a significance of $\sim$7$\sigma$ 
at a position corresponding to R.A.(2000)=22h51m42.72s
and Dec(2000)=+22$^{\circ}$17'56.4" and with a positional uncertainty
of  4.5' (90\% confidence level). Within the INTEGRAL/IBIS error box we find the high redshift QSO MG3 J225155+2217 (z=3.668) 
but only follow up observations with the Swift satellite  using the XRT telescope  could confirm  
this association with certainty.
Indeed only one bright and hard X-ray source was detected within the IBIS positional uncertainty and this source is the high z quasar (see left panel in  Figure \ref{qso}).
A simple power law fit to the IBIS data provides a
good fit ($\chi^2$=6.5 for 8 d.o.f.)
and a photon index $\Gamma$=1.4$\pm$0.6 combined to an observed  20-100 keV (around 100-500 keV in the source rest frame)
flux of 4 $\times$ 10$^{-11}$ erg cm$^{-2}$ s$^{-1}$.

\begin{figure*}[t]
\begin{minipage}{8cm}
 \includegraphics[height=0.35\textheight]{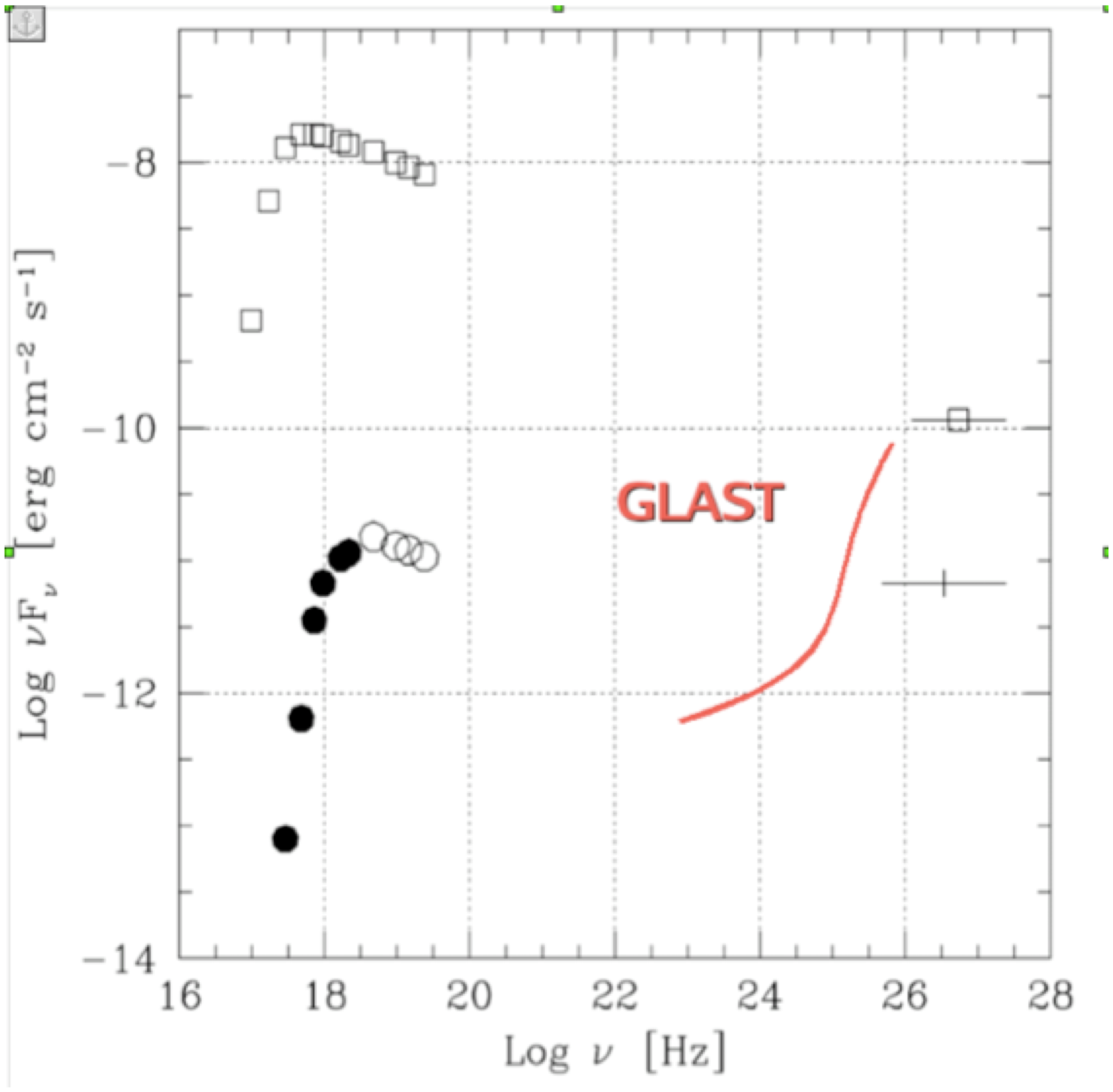}
\end{minipage}
\hspace{1cm}
\begin{minipage}{8cm}
	 \includegraphics[height=0.3\textheight]{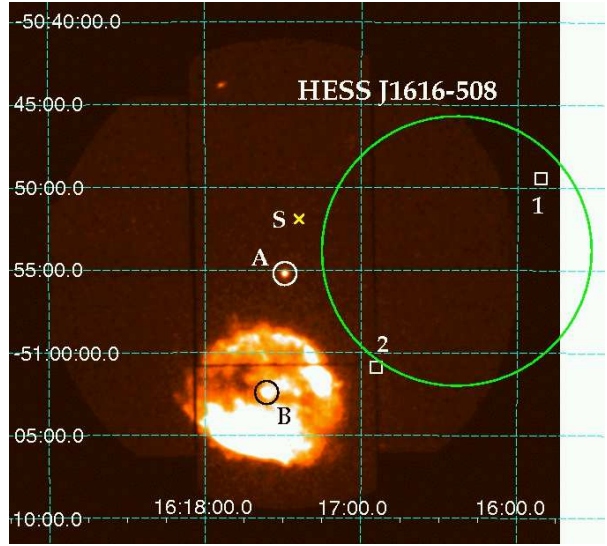}
\end{minipage}
 \caption{Left: Spectral energy distribution of IGR J18135-1751=HESS
  J1813-178 (circles) and The SED of Crab nebula is also plotted
   (open squares). GLAST sensitivity is plotted in red. Right: XMM-Newton MOS 0.5--10 keV image of the region around the HESS J1616--508.
PSR J1617--5055 and 1E 161348--5055.1 (circles labelled as A and B respectively) 
are the only sources detected at energies above 4 keV.
We also plot the position (boxes labelled as 1 and 2)
of the two sources detected within the XRT field of view.
The source labelled as S (yellow cross) is associated to the star HD 146184.}
\label{hess1}
\end{figure*}

A power law  also provides a good fit to the Swift/XRT data and consistent results in spectral shape for the two longest observations availbale in the archive
($\Gamma$=1.41$\pm$0.12 versus $\Gamma$=1.30$\pm$0.08 ); the flux was instead 
a slightly different 2-10 keV flux and if we also use a third shorter observation to estimate only the source flux, 
a trend of decreasing 
flux from 4.4 to 2.7 $\times$ 10$^{-12}$ erg cm$^{-2}$s$^{-1}$ becomes evident over a 6 day period.
To improve the statistics and in view of the fact that the  INTEGRAL detection is over a few revolutions,
i.e provides an average flux,
we have combined the two most  statistically significant XRT spectra and repeated the analysis.
This average spectrum has a flat photon index ($\Gamma$=1.31$\pm$0.09 ) and an acceptable 
$\chi^{2}$ =46.9/53. However, this  model results in residuals that show some  curvature possibly due to intrinsic 
absorption in the source rest frame.  Addition of this extra component  provides a fit improvement which is significative 
at the 99.98 $\%$ confidence level according to the F test, a  more typical AGN spectrum
( $\Gamma$= 1.53$\pm$0.16 ) and a mild column  density (N$_{H}$=3$\pm$2 $\times$ 10$^{22}$ cm$^{-2}$).
However, since absorption cannot be the only cause of a deficit of soft photons  in high redshift QSO \cite{tavecchio07}, 
we have also tried a broken power law:
this model provides an equally significant improvement in the fit, a break energy at 1.55$\pm$ 0.15 keV and a spectral flattening
below 1-2 keV of $\Delta \Gamma$= 0.5 ($\Gamma_{1}$=0.7$\pm$0.3 and $\Gamma_{2}$=1.2$\pm$0.4).  
The broad band Swift/XRT and INTEGRAl/IBIS spectrum fitted with an absorbed power law fit is shown in the right panel in Figure \ref{qso}:
while we find a perfect match in spectral shape with $\Gamma$=1.5, the XRT data fall short of the INTEGRAL detection 
by a factor in the range 3-7, implying again variability 
in the source flux as also evident in the sequence of XRT observations.

Assuming that the IBIS observation  represents the average state of the  source, we obtain rest frame luminosities\footnote  {We adopt H$_{o}$=71 km s$^{-1}$ Mpc$^{-1}$, $\Omega_{\Lambda}$=0.73 and $\Omega_{M}$=0.27}
of 0.3$\times$10$^{48}$ erg  s$^{-1}$ in the X-ray (2-10 keV) 
band,  2$\times$10$^{48}$ erg  s$^{-1}$ at  hard X-rays  (20-100 keV) and 
5$\times$10$^{48}$ erg s$^{-1}$ in the soft gamma-ray  (100-500 KeV)
interval, i.e. MG3 J225155+2217 is an X/gamma-ray lighthouse shining from  the edge of our Universe.
Further analysis of the source characteristics indicates that IGR J2251+2218 is a blazar 
belonging to the class of Flat Spectrum Radio QSO. 
The source SED is also unusual among  such type of blazars as it is compatible with having 
the synchrotron peak in the X/gamma-ray band ( i.e. much higher than 
generally observed) or alternatevely with the Compton peak in the MeV range 
(i.e. lower than  typically measured). Indeed,
the  X-ray to radio flux ratio is $\sim$ 1000 (or $\alpha_{xr}$ 
$<$ 0.75), i.e. similar to high energy peaked blazars, i.e. those 
with the synchrotron peak in the X-ray band; this is further supported by the observed
shape of the infrared to optical continuum which is at odds with the 
location of  the synchrotron peak  at infrared frequencies as generally 
observed in FSRQ. 
Obviously, the sparse and non-simultaneous data coverage still leaves open 
the possibility of a more classical double humped interpretation of the 
source SED; but even in this case the object is strange as not only the 
observed peculiarities have to be explained, but in this case the Compton peak 
would be at MeV energies.
Either way, MG3 J225155+2217 is quite atypical for its class and an extreme object in the blazar population;
this makes it an interesting laboratory in where to test current blazars theories and so an object worth following up at all wavebands.

\section{The hard-X/soft-Gamma emission from HESS sources}
\label{hess}

\begin{figure*}[t]
\begin{minipage}{8cm}
\includegraphics[height=0.3\textheight]{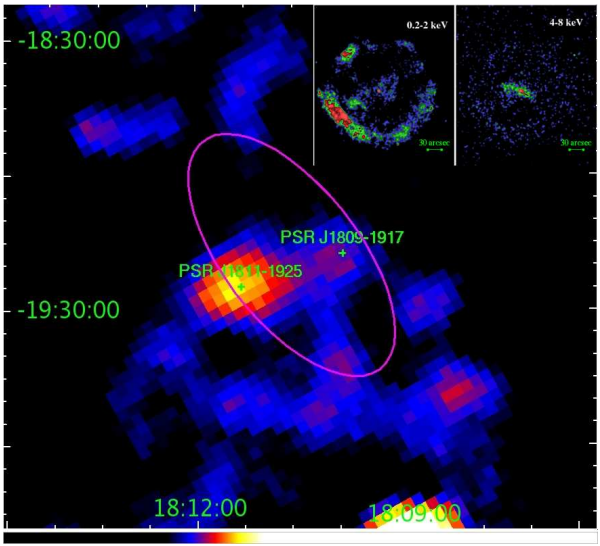}
\end{minipage}
\hspace{1cm}
\begin{minipage}{8cm}
\includegraphics[width=0.85\linewidth, angle=-90]{composite_spec_all.ps}
\end{minipage}
\caption{Left: integral image of the region around PSR J1811-1925. The HESS extension is reported with a magenta ellipse. In the inset we sho the chandra images in 0.2--2 keV (on the left) and 4--8 keV (on the right). Right: Composite spectrum chandra and integral. The two data set in the chandra energy range are the spectra extractet from the jet-like feature (bottom) and PST+jet-like feature (upper)}
\label{hess2}
\end{figure*}

The first sensitive TeV surveys  of the inner part of our Galaxy by HESS (High Energy Steroscopic System) collaboration \cite{aharonian}, 
revealed the existence of a population of high energy Gamma-ray
objects, several of which previously unknown or not yet identified at
lower wavelength. 

Isolated pulsars/pulsar wind nebulae (PWN), supernova
remnants (SNR), star forming regions, binary systems with a collapsed
object like a microquasar or a pulsar, can be all cosmic particle accelerators and potential sources of high energy Gamma ray. 
To discriminate between  various emitting scenarios and, in turn, to fully understand
the nature of these Very High Energy (VHE) sources, detection in hard-X and soft-Gamma ray are foundamental. The IBIS Gamma-ray imager on board INTEGRAL is a
powerful tool to search for their counterpart above 20 keV in view of
the arcmin  Point Source Location Accuracy associated to $~$ millicrab
sensitivity for exposure $>$1 Ms.
We described in the following successfull association of some VHE emitter with INTEGRAL source.
However in some cases is not possible to firmly associate the TeV emission to the objects detected by INTEGRAL.
In fact, even if the energetic budget in a standard emmiting scenario (Self Synchrotron Compton+Inverse Compton) goes in the right direction for the asscociation, 
often the morphology of the INTEGRAL source and the crowded field do not allow to reach a firm conclusion.
HESS J1813-178 has a point like X-ray counterpart, IGR J18135-1751 \cite{ube05}, with a power law emission  from 2
to 100 keV and an associated radio counterpart. It is a non-thermal source,
possibly accelerating electrons and positrons which radiate through
synchrotron and inverse Compton mechanism. This is suggestive of
the presence of a PWN/SNR, as already found in most newly detected TeV
objects \cite{aharonian} that have been clearly associated with either
shell-type or plerion-type supernova remnants, like 
AX J1838-0655/HESS J1837-069 \cite{malizia05}.
The lack of strong X/Gamma-ray variability in IGR J1813-178 as well as AX
J1838-0655, makes unlikely the scenario in which the TeV emission is
due to a binary system with a pulsar as compact object as observed in
HESS J1303-631/PSR B1259-63 \cite{aharonian2}.
In the left panel in Figure \ref{hess1} we plot the SED of IGR J1813-178
together with the SED of Crab nebula, and the GLAST-LAT sensitivity curve for 1 year of exposure.
The shape of the
SED is similar although with a quite different ratio X-ray to TeV
Gamma ray. Also for PWNs, the ratio between the luminosity in X-ray
and that in radio is still an open issue: there
are sources of this type with X-ray luminosities similar to the Crab but
with  radio fluxes 2 to 3 order of magnitude weaker \cite{slane}.
Finally the observed X-ray, soft Gamma and TeV luminosities, 4, 3.4, (1.2-1.9)$\times 10^{34}$ erg/s, 
 are similar to the values observed
in the few HESS sources which have been clearly identified with PWNs
or shell type SNRs.

HESS J1616--508 is one of the brightest emitters in the TeV sky. Recent observations with the
IBIS/ISGRI have revealed that a young nearby
and energetic pulsar, PSR J1617--5055, is a powerful emitter of soft Gamma-ray in the 
20--100 keV domain. The analysis of all available data from the 
{INTEGRAL, Swift, BeppoSAX and XMM-Newton telescopes show evidence
 that the energy source that fuels the 
X/Gamma-ray emissions is very likely derived from the pulsar, both on the basis of the 
positional morphology and the energetics of the system \cite{landi07}. We show  in the right panel in Figure \ref{hess1} XMM-Newton observation of PSR J1617--5055.  Likewise the 1.2$\%$ of the pulsar's spin 
down energy loss needed to power the 0.1--10 TeV emission is also fully consistent with other
HESS sources known to be associated with pulsars.
The observed 2--10 and 20--100 keV fluxes of $4.2 \times10^{-12}$ erg cm$^{-2}$ s$^{-1}$
and $1.37 \times10^{-11}$ erg cm$^{-2}$ s$^{-1}$ translate, at the distance of
the pulsar, to luminosities which are approximately $\sim 2\times10^{34}$ erg s$^{-1}$
and $\sim 7\times10^{34}$ erg s$^{-1}$, or 0.1\% and 0.4$\%$ of
the spin down losses. These values are fully within	
the range observed from the pulsar wind nebulae systems observed by
INTEGRAL.
The TeV emission is not positionally coincident with the
pulsar or any other nearby object except for two X-ray sources detected by
Swift/XRT. However, the lack of precise positional coincidence between HESS J1616--508
and PSR J1617--5055 does not necessarily
dissociate the two objects, particularly in view of recent observational
evidence of TeV sources offset from their powering pulsars

However it is not easy to obtain direct associations between VHE emitters and INTEGRAL sources. 
In the particular case of PSR~J1811--1925 the composite X-ray by Chandra and soft gamma-ray spectrum 
by INTEGRAL (see left panel in Figure \ref{hess2})  indicates that the
pulsar provides around half of the emission seen in the soft gamma-ray domain; its spectrum
is hard with no sign of a cut off up to at least 80 keV (see rigt panel Figure \ref{hess2}). The other half of the emission above
10 keV comes from the PWN; with a $\Gamma$=1.7 its spectrum is softer than that of the pulsar.
This is one of the few cases (notably Crab and Vela) where emission above 10 keV is detected from a PWN.
The nature of this hard X-ray emission is not yet clear, as both
the synchrotron and inverse Compton processes are likely emission
mechanisms.
In the case of PSR~J1811--1925 the association with HESS~1809--193 is highly improbable. The PSR~J1811--1925 is the most energetic yung pulsar within the region surronding the TeV source, but not the only one capable of providing the requisite energy. PSR~J1809--1917 provides a more plausible counterpart, although it is still possible that neither fuels the VHE emitter, which could instead be powered by one of the supernova remnants which exist in the region \cite{dean07}.

As shown in Figure \ref{hess1}, left panel, with its high sensitivity, GLAST will provide spectral 
measurements in an energy regime in which differences between hadronic and leptonic model  for
Gamma-ray production are significant \cite{funk07}, in turn pinpointing the emission mechanism taking place in this new class 
of very high energy emitters. In additon AGILE observations in the GeV energy region, together with 
INTEGRAL and soft X-rays observations (e.g. with the great spacial resolution available with Chandra), will allow us to build a 
detailed SED of these VHE sources in order to provide the radiative mechanism that produces the observed  emssion.

\section{Conclusions}
\label{conclusions}
After 5 years of observations, INTEGRAL is performing a survey of the all sky and the deep monitoring of the Galactic Plane, observing more tha 420 sources; this is giving us exiting results about new high energy emitters and discovering new dozen of variable sources.
The large field of view, the good angular resolution and the deep observations make all together a powerfull tool to discovery high variable sources, like new SFXTs, distant AGN, like IGR J22517+2218 the second most distant blazar detected above 20 keV and a gamma-ray lighthouse shining from the edge of our universe, and the X-rays counterpart at the VHE sources detected in Gamma ray.\\

The authors aknowledge financial contribution from contract ASI-INAF I/023/05/0 and ASI I/008/07/0


\begin{thebibliography}{00}


\bibitem{aharonian} 
Aharonian F., et al. 2005, {\em Science} {\bf 307} 1938

\bibitem{aharonian2} 
Aharonian F.,  et al. 2005b {\em A\&A} {\bf 442} 1

\bibitem{bassani07} 
Bassani, L. et al.  2007 Ap.J. 669,L1

\bibitem{bazzano06}
Bazzano A., et al., 2006 ApJ, 649, L9

\bibitem{bird07} 
Bird A.J. et al., 2007, 2007, ApJ Supplement, 170, 175

\bibitem{cappi06} 
Cappi M. et al. 2006 A.\&A 446, 459

\bibitem{chaty06} 
Chaty, S., Rahoui, F., 2006, astro-ph/0609474 

\bibitem{courv03}
Courvoisier  T., et al. 2003, A\&A, 411 L49 

\bibitem{dean07}
Dean A. J., et al. 2008, 384, MNRAS, L29


\bibitem{disalvo06}
Di Salvo T., et al. 2006, ApJL, 649, L91

\bibitem{filippenko89}
Filippenko A. V. \&  Sargent W.L.W. 1989, Ap.J. 342, L11

\bibitem{funk07} 
Funk S.  2007, in Advances in Space Research (Proceedings COSPAR 2006), arXiv:astro-ph/0701471v1

\bibitem{gehrels04}
Gehrels, N., et al. 2004, Ap.J. 611, 1005
\bibitem{hill04}
Hill, J. E., et al. 2004, Proc. SPIE 5165, 21


\bibitem{intzand05} 
In't Zand J. 2005,  A\&A, 441, L1

\bibitem{landi07} 
Landi R, et al. 2007, MNRAS, 380, 926

\bibitem{lund03}
Lund N.,  et al. 2003, A\&A, 411, L231

\bibitem{malizia05}
Malizia A., et al. 2005, ApJL, 630, 157

\bibitem{masetti06}
Masetti N., et al. 2006, A\&A, 459, 21

\bibitem{mass03}
Mass-Hesse M. et al. 2003 A\&A , 411 L261

\bibitem{moran05}
Moran, E.C., et al.  2005 A.J. 129, 2108

\bibitem{negu05} 
Negueruela, I., Smith, D. M., Reig, P., et al. 2005, ESA SP-604, 165

\bibitem{negu06} 
Negueruela, I., Smith, D. M., Harrison T. E., 2006, ApJ, 638, 982 

\bibitem{panessa06}
Panessa, F. et al.  2006 A.\&A. 455, 173

\bibitem{sguera06}
 Sguera, V., Bazzano, A., Bird, A.J., et al., 2006, ApJ, 646, 452

\bibitem{sguera05}
 Sguera, V., Barlow, E. J., Bird, A. J.,  et al.,  2005, A\&A, 444, 221

\bibitem{sidoli06} 
Sidoli, L., Paizis, A., Mereghetti, S., 2006, A\&A, 450L, 9S 

\bibitem{slane} 
Slane, P.,  et al. 2001, {\em ApJ} {\bf 548},  814

\bibitem{skiner03}
Skinner, G. \& P.Connell 2003, A\&A, 411, L123

\bibitem{smith06} 
Smith, D. M., Heindl, W. A., Markwardt, C. B., et al. 2006, ApJ, 638, 974

\bibitem{tavecchio07} 
Tavecchio, F., et al.  2007, Ap.J. 665, 980

\bibitem{ube03} 
Ubertini P., Lebrun, F., Di Cocco, G., et al. 2003, A\&A, 411, L131

\bibitem{ube05} 
Ubertini P., et al. 2005, ApJL, 29, 109

\bibitem{vaughan05} 
Vaughan, S., et al. 2005 MNRAS 356, 524

\bibitem{vedrenne03}
Vedrenne G,. et al. 2003 A\&A, 411, L63  

\bibitem{walter06} 
Walter, R., Zurita Heras, J., Bassani, L., et al. 2006, A\&A, 453, 133

\bibitem{winkler03}
Winkler C.  et al. 2003, A\&A, 411, L1 



\end{thebibliography}
\end{document}